\documentclass{aa}  

\usepackage{graphicx}
\usepackage{subfigure}
\usepackage{txfonts}
\usepackage{xcolor}
\usepackage{tablefootnote}
\usepackage{color,soul}

\begin{document} 

\title{Constraining bright optical counterparts of fast radio bursts}
\titlerunning{Constraining bright optical counterparts of FRBs}

\author{Consuelo N\'u\~nez\inst{1},
Nicolas Tejos\inst{1},
Giuliano Pignata\inst{2,3},
Charles D. Kilpatrick\inst{4},
J. Xavier Prochaska\inst{5,6},
Kasper~E.~Heintz\inst{7,8,9},
Keith W. Bannister\inst{10},
S. Bhandari\inst{10},
Cherie K. Day\inst{11,10},
A.~T.~Deller\inst{11},
Chris Flynn\inst{11,12},
Elizabeth K. Mahony\inst{10},
Diego Majewski\inst{5},
Lachlan Marnoch\inst{13,10,14},
Hao Qiu\inst{15},
Stuart D. Ryder\inst{13,14},
Ryan~M.~Shannon\inst{11}
}

\institute{
Instituto de F\'isica, Pontificia Universidad Cat\'olica de Valpara\'iso, Casilla 4059, Valpara\'iso, Chile\\
\email{consuelo.nunez.p@mail.pucv.cl; nicolas.tejos@pucv.cl}
\and
Departamento de Ciencias F\'isicas, Universidad Andres Bello, Avda.  Republica 252, Santiago, Chile
\and
Millennium Institute of Astrophysics, Nuncio Monse\~nor S\'otero Sanz 100, Providencia, Santiago, Chile
\and
Center for Interdisciplinary Exploration and Research in Astrophysics (CIERA) and Department of Physics and Astronomy, Northwestern University, Evanston, IL 60208, USA
\and
University of California Observatories-Lick Observatory, University of California, 1156 High Street, Santa Cruz, CA95064, USA
\and
Kavli Institute for the Physics and Mathematics of the Universe (WIP), 5-1-5 Kashiwanoha, Kashiwa, 277-8583, Japan
\and
Centre for Astrophysics and Cosmology, Science Institute, University of Iceland, Dunhagi 5, 107 Reykjav\'ik, Iceland
\and
Cosmic Dawn Center (DAWN), Denmark,
\and
Niels Bohr Institute, University of Copenhagen, Jagtvej 128, 2200 Copenhagen N, Denmark
\and 
Australia Telescope National Facility, CSIRO Astronomy and Space Science, PO Box 76, Epping, NSW 1710, Australia
\and
Centre for Astrophysics and Supercomputing, Swinburne University of Technology, Hawthorn, VIC 3122, Australia
\and
ARC Centre of Excellence for Gravitational Wave Discovery (OzGrav), Australia
\and
Department of Physics and Astronomy, Macquarie University, NSW 2109, Australia
\and
Astronomy, Astrophysics and Astrophotonics Research Centre, Macquarie University, Sydney, NSW 2109, Australia
\and
Sydney Institute for Astronomy, School of Physics, University of Sydney, NSW 2006, Australia
}

\authorrunning{C.~N\'u\~nez et al.} 


 
\abstract{Fast radio bursts (FRBs) are extremely energetic pulses of millisecond duration and unknown origin. 
To understand the phenomenon that emits these pulses, targeted and un-targeted searches have been performed for multiwavelength counterparts, including the optical.}
  {The objective of this work is to search for optical transients at the positions of eight well-localized ($<1 \arcsec$) FRBs after the arrival of the burst on different timescales (typically at one day, several months, and one year after FRB detection). We then compare this with known optical light curves to constrain progenitor models.}
  {We used the Las Cumbres Observatory Global Telescope (LCOGT) network to promptly take images 
  with its network of 23 telescopes working around the world. We used a template subtraction technique 
  to analyze all the images collected at differing epochs. 
  We have divided the difference images into two groups: In one group we use the image of the last epoch as a template, and in the other group we use the image of the first epoch as a template. 
  We then searched for optical transients at the localizations of the FRBs in the template subtracted images.}
  {We have found no optical transients and have therefore set limiting magnitudes to the optical counterparts. Typical limits in apparent and absolute magnitudes for our LCOGT data are $\sim 22$ and $-19$\,mag in the $r$ band, respectively. We have compared our limiting magnitudes with light curves of super-luminous supernovae (SLSNe), Type~Ia supernovae (SNe~Ia), supernovae associated with gamma-ray bursts (GRB-SNe), a kilonova, and tidal disruption events (TDEs).}
  {Assuming that the FRB emission coincides with the time of explosion of these transients, we rule out associations with SLSNe (at the $\sim99.9\%$ confidence level) and the brightest subtypes of SNe~Ia, GRB-SNe, and TDEs (at a similar confidence level). However, we cannot exclude scenarios where FRBs are directly associated with the faintest of these subtypes or with kilonovae.}

   \keywords{fast radio burst --
                supernovae: general --
                techniques: photometric
               }

   \maketitle

\begin{table*}[t]
\caption{LCOGT observation summary.}             
\label{tab:observations}      
\centering                          
\begin{tabular}{c c c c c c}        
\hline\hline                 
Field & Date & Site\tablefootmark{a} & Seeing  & Moon  & Background noise  \\
& (UTC) & & (\arcsec) & illumination & RMS (counts) \\
\hline                        
FRB180924 & 2019-05-31 15:50:34 & SSO & $2.0$ & $15\%$ & $4.0$\\
          & 2020-06-29 17:28:15 & SSO & $1.6$ & $53\%$ & $3.9$\\
\hline
FRB181112 & 2019-05-31 19:27:09 & SSO  & $1.5$ & $15\%$ & $4.5$\\
          & 2020-06-29 17:39:45 & SSO  & $3.5$ & $53\%$ & $4.4$\\
\hline
FRB190102 & 2019-05-31 17:32:31 & SSO & $2.3$ & $15\%$ & $4.4$\\
          & 2020-06-29 17:44:52 & SSO & $2.0$ & $53\%$ & $4.3$\\
\hline
FRB190608 & 2019-06-09 03:41:26 & SAAO & $2.2$ & $35\%$ & $4.6$\\
          & 2019-08-08 14:54:55 & SSO & $2.2$ & $48\%$ & $4.8$\\
          & 2020-06-29 17:23:27 & SSO & $3.3$ & $53\%$ & $4.7$\\
\hline
FRB190611 & 2019-06-11 17:31:37 & SSO & $1.9$ & $54\%$ & $4.6$\\
          & 2019-08-08 18:01:28 & SSO & $2.4$ & $48\%$ & $4.7$\\
          & 2020-06-29 17:55:35 & SSO & $3.4$ & $53\%$ & $4.9$\\
\hline
FRB190711 & 2019-07-11 03:45:05 & SAAO & $2.3$ & $66\%$ & $4.3$\\
          & 2019-08-09 03:46:15 & CTIO & $2.3$ & $54\%$ & $6.7$\\
          & 2020-06-29 17:12:12 & SSO  & $1.8$ & $53\%$ & $4.1$\\
\hline
FRB190714 & 2019-07-14 17:03:40 & SAAO & $1.7$ & $91\%$ & $8.8$\\
          & 2019-08-08 23:10:34 & CTIO & $3.5$ & $54\%$ & $13.1$\\
          & 2020-06-30 08:11:14 & SSO & $1.6$ & $53\%$ & $12.2$\\
\hline
FRB191001 & 2019-10-05 20:17:22 & SAAO & $2.1$ & $42\%$ & $4.5$\\
          & 2019-11-26 00:53:07 & CTIO & $2.3$ & $0\%$ & $4.9$\\
          & 2020-06-29 17:06:57 & SSO & $3.4$ & $53\%$ & $3.9$\\
\hline  
\end{tabular}
\tablefoot{
\tablefoottext{a}{SSO: Siding Spring Observatory. SAAO: South African Astronomical Observatory. CTIO: Cerro Tololo Inter-American Observatory.}
}
\end{table*}

\section{Introduction}

Fast radio bursts (FRBs) are extremely energetic radio frequency pulses that last for milliseconds or less \citep[see, e.g.,][for a review]{cordes}. The dispersion measure (DM) of FRBs is greater than the expected contribution of the Milky Way \citep[e.g.,][]{petroff}, which implies they are extragalactic. These signals come from all directions in the sky, and a sky rate of [$818\pm64$(stat.)$_{-200}^{+220}$(sys.)]\,sky$^{-1}$\,day$^{-1}$ above a fluence of $5$\,Jy~ms at $600$\,MHz  has been estimated \citep{cat_chime}.

Hundreds of FRBs have been reported thus far by different radio telescopes \citep[e.g.,][]{lorimer,spitler14,masui15,caleb17,shannon18,bhandari18,fedorova19,ravi19,macquart19,law20,connor20,cat_chime}, but only a dozen of them are localized to subarcsecond precision \citep{121102, localized, marcote}. For these few sources it has been possible to identify their host galaxies and redshifts, which confirms that they come from extragalactic sources.  Most of these localizations, including those involved in this paper, have been detected by  the Australian Square Kilometre Array Pathfinder (ASKAP) telescope \citep{askap}.

The first FRB was discovered in 2007 by \citet{lorimer}, and since then identifying the physical phenomenon (or phenomena)
that gives rise to these bursts has been pursued. Many theories have been proposed for possible progenitors, including some kinds of supernovae (SNe; see below), compact-object mergers involving neutron stars (NSs), white dwarfs (WDs), and/or black holes (BHs), among many others \citep[see, e.g.,][for compilations]{platts, chatterjee}. The recent detection of an intense radio burst within the Milky Way from the magnetar SGR~1935+2154 hints that at least part of the FRB population originates from magnetars \citep{magnetar, magnetar2, magnetar3}. 

Regarding SNe, some scenarios involve core-collapse supernovae (CCSNe), super-luminous supernovae (SLSNe) associated with long gamma-ray bursts (LGRBs), and Type Ia supernovae \citep[SNe~Ia;][]{kashiyama13, connor16, metzger17}. Young magnetars or pulsars immersed in SN remnants \citep[see, e.g.,][]{michilli} can explain the observed DMs and rotation measures (RMs). There is also the possibility that FRBs originate from WD mergers \citep{kashiyama13}, which would produce an SN Ia. However, these models predict that it would take tens to hundreds of years for the SN ejecta to dissipate enough for FRB pulses to penetrate it \citep[e.g.,][]{piro}.

\begin{table*}[t]
\caption{FRB fields and epochs of observations.}             
\label{tab:epochs}      
\centering                          
\begin{tabular}{c c c c c c c c}        
\hline\hline                 
FRB & Arrival Time & $1^{st}$ epoch & $2^{nd}$ epoch & $3^{rd}$ epoch & $z$ & Distance & Galactic\\
& (UTC) & (days) & (days) & (days) & & Modulus\tablefootmark{a} & Extinction\tablefootmark{b} \\
\hline                        
FRB180924 & 2018-09-24 16:23:12 & 249 & 644 & - & 0.3212 & 41.15 & 0.04\\
FRB181112 & 2018-11-12 17:31:15 & 200 & 595 & - & 0.4755 & 42.16 & 0.05\\
FRB190102 & 2019-01-02 05:38:43 & 149 & 545 & - & 0.2912 & 40.91 & 0.52\\
FRB190608 & 2019-06-08 22:48:13 & 0.20 & 60.7 & 387 & 0.1177 & 38.72 & 0.11\\
FRB190611 & 2019-06-11 05:45:43 & 0.49 & 58.5 & 385 & 0.3778 & 41.57 & 0.52\\
FRB190711 & 2019-07-11 01:53:41 & 0.08 & 29.1 & 355 & 0.5220 & 42.41 & 0.32\\
FRB190714 & 2019-07-14 05:37:13 & 0.48 & 25.7 & 352 & 0.2365 & 40.39 & 0.14\\
FRB191001 & 2019-10-01 16:55:36 & 4.14 & 55.3 & 272 & 0.2340 & 40.37 & 0.07\\
\hline                                   

\end{tabular}
\tablefoot{
\tablefoottext{a}{Using $\Lambda$ cold dark matter cosmological parameter from \cite{WMAP9}.}
\tablefoottext{b}{Galactic extinction, $A_r$, based on \cite{extinction}.}
}
\end{table*}

To further examine the origin of FRBs, counterparts at different wavelengths have been sought, for example in the optical range \citep{optical, op_follow_up, marnoch, optical2}, X-rays \citep{multiwavelength2, multiwavelength, xray, xray2, xray3}, and gamma rays \citep{gamma, gamma2, gamma3, gamma4}. However, most of these searches have been reactive; that is, first the FRB is detected in the radio and then observations are triggered at the different wavelengths (\citealt{simultaneous}). This leads to a considerable time delay for these multiwavelength follow-up observations. 

In \cite{marnoch} the first three FRBs well localized by ASKAP were used to search for SN-like transient optical counterparts with the Very Large Telescope (VLT). They triggered one image $10-46$ days after the burst detection and one image $233-333$ days later
to serve as a template for image differencing.
They found no optical counterpart, so they put limits on the brightness of potential transients. They modeled light curves of different types of SNe, concluding that SNe~Ia~and Type IIn supernovae (IIn SNe) are unlikely to be associated with any non-repeating FRBs.

Recently, \cite{optical2} performed optical follow-up of FRB180916 on 30 second timescales (with a time delay of seconds to minutes) with the Apache Point Observatory (APO) to constrain the presence of optical emission contemporaneous with a radio burst. The repeating FRB180916 has a well-established period of $\sim16.3$ days, which has made a coordination of observations with the Canadian Hydrogen Intensity Mapping Experiment (CHIME) possible in the radio. While CHIME did detect a radio pulse, no optical transient was apparent within a few seconds of the FRB arrival to a depth of $r\approx24.5$\,mag. From these limits, \cite{optical2} ruled out a synchrotron maser from repeating magnetar flares where the burst energy was $> 10^{44}$ erg and the circumburst density was $> 10^{4}$ cm$^{-3}$.

In this work we present a search for optical transients using the Las Cumbres Observatory Global Telescope (LCOGT) network \citep{brown} at the positions of eight well-localized FRBs detected by ASKAP \citep{localized, heintz}, with data from the day of arrival of the FRB and up to $\sim1-2$ years later. 
We identify no transients in our search and therefore present limits on emission for classes of luminous optical transients.

Compared to previous optical follow-up work, here we analyze a sample of several FRBs rather than single events \citep[][but see \citealt{marnoch}]{optical, op_follow_up, optical2}. In addition, we set limits on days very close to the arrival of the bursts ($\sim 1$\,day), and our limits extend to $\sim1$\,year after the FRB emission.

Despite the modest aperture of our telescope network, we were able to constrain several extreme but possible optical transient scenarios, such as SLSNe and the brightest subtypes of SNe~Ia, supernovae associated with gamma-ray bursts (GRB-SNe), and tidal disruption events \citep[TDEs; e.g.,][]{TDE}, assuming that the FRB emission coincides with the time of explosion of these transients. 
While TDEs are disfavored as a dominant channel
based on studies of FRB offset from the host galaxy centers \citep[e.g.,][]{localized,heintz,mannings}, here we test this scenario based solely on their optical light curves.

Our paper is structured as follows. In Sect.~\ref{sec:data} we present our observation strategy and the data obtained with LCOGT. In Sect.~\ref{sec:analysis} we search for optical transients and place limiting magnitudes based our observations. The results are presented in Sect.~\ref{sec:results}, and the conclusions are presented in Sect.~\ref{sec:conclusions}.

\section{Data}
\label{sec:data}

\subsection{LCOGT images}

 We used data from the  LCOGT network \citep{brown} for this analysis. LCOGT is a network of 23 telescopes at seven sites around the world. We used the 1m telescopes with Sinistro cameras, which have a field of view (FoV) of $26.5 \arcmin \times 26.5 \arcmin$ and a pixel size of $0.389 \arcsec$. Dates, sites, seeing, Moon illumination, and background noise RMS of all observations are listed in Table \ref{tab:observations}. The images were automatically processed through the BANZAI pipeline \citep{banzai} with the latest calibration frames\footnote{i.e., reduction level code 91.}. 

The typical limiting apparent magnitudes obtained with our LCOGT data are approximately $22$\,mag in the $r$ band (see Sect.~\ref{sec:lower_limits}). Nevertheless, one great advantage of using LCOGT for transient follow-up is that the network is composed of telescopes all around the world, which allowed us to obtain images of each FRB field 
promptly ($<1$\,day for most of the fields).

\subsection{Observation strategy}

Our observation strategy was to manually take an initial ``epoch'' of optical imaging (ideally as soon as the FRB is detected) followed by an additional epoch $30-60$\,days after the radio burst to search for transient emission. For each epoch we obtained a set of ten images of $60$\,s in the $r$ band. 

We triggered observations from eight FRB fields, four of which were followed-up on with rapid response, which means the first epoch could be taken within the first day after the FRB detection; however, the fields FRB180924, FRB181112, and FRB190102 have first epochs taken with much longer delays ($\sim 100$\,days). We still included them in the analysis as these can constrain longer timescale transients such as SLSNe. Table~\ref{tab:epochs} summarizes the time delays incurred in each epoch for our different fields. 

In the first epochs of FRB190608, FRB190611, FRB190711, and FRB190714, the images were taken using localizations with larger positional uncertainties ($\approx 20 \arcmin \times 20 \arcmin$) as given by the analysis from ASKAP based on incoherent sum data \citep[e.g.,][]{shannon18}. For their second and third epochs, we triggered the observations centered on the much more accurate position ($\sim 1-2$\arcsec) as given by the subsequent coherent ASKAP analysis \citep{bannister19}. 

From this data set we can make multiple comparisons between different epochs to search for putative bright optical transients. If no transient is seen, we set limiting magnitudes (upper limits in fluxes).

\begin{figure*}[t]
    \centering
    \subfigure[FRB190608 in $1^{st}$ epoch.]
    {\includegraphics[width=0.3\textwidth]{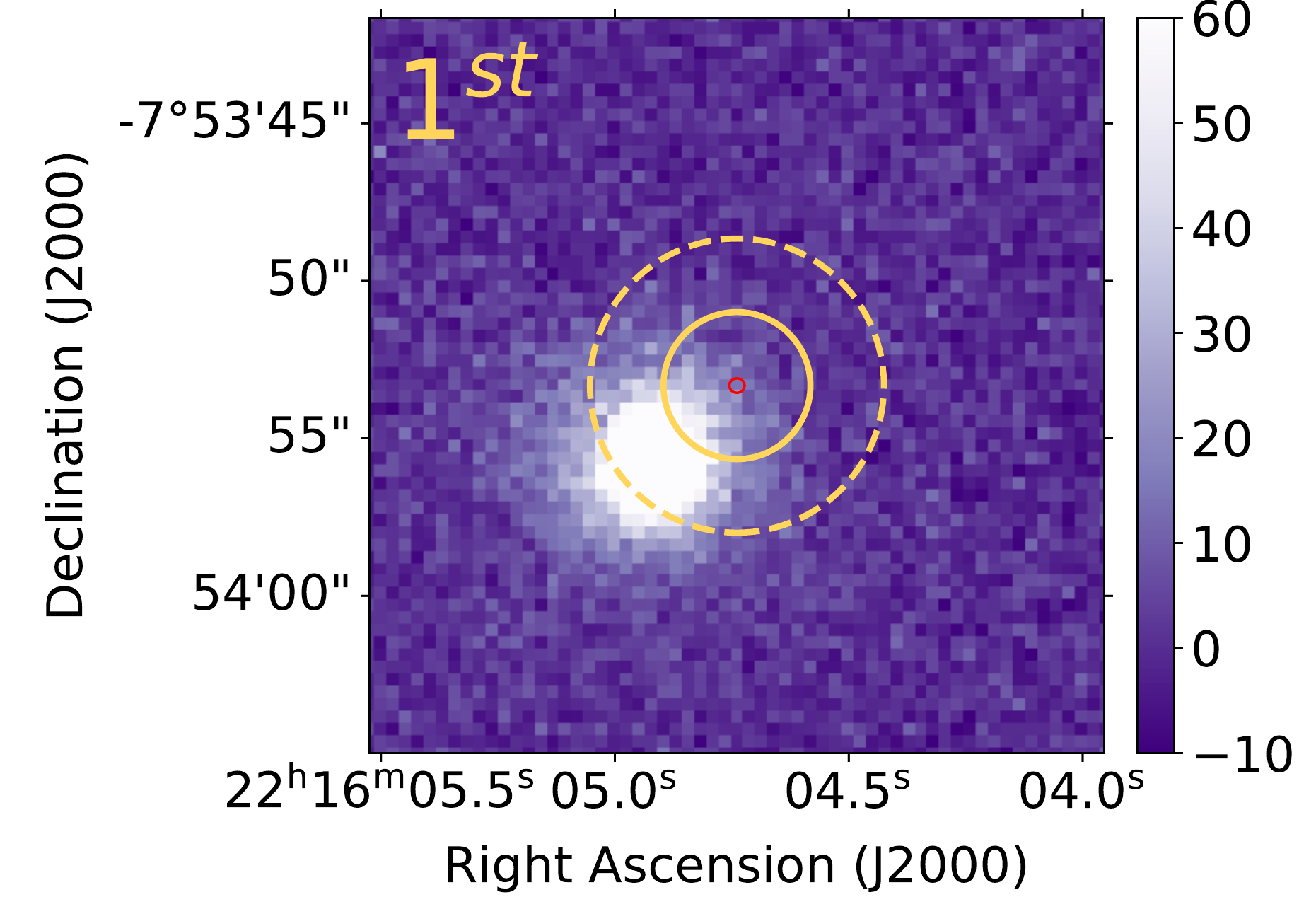}}
    \subfigure[FRB190608 in $2^{nd}$ epoch.]
    {\includegraphics[width=0.3\textwidth]{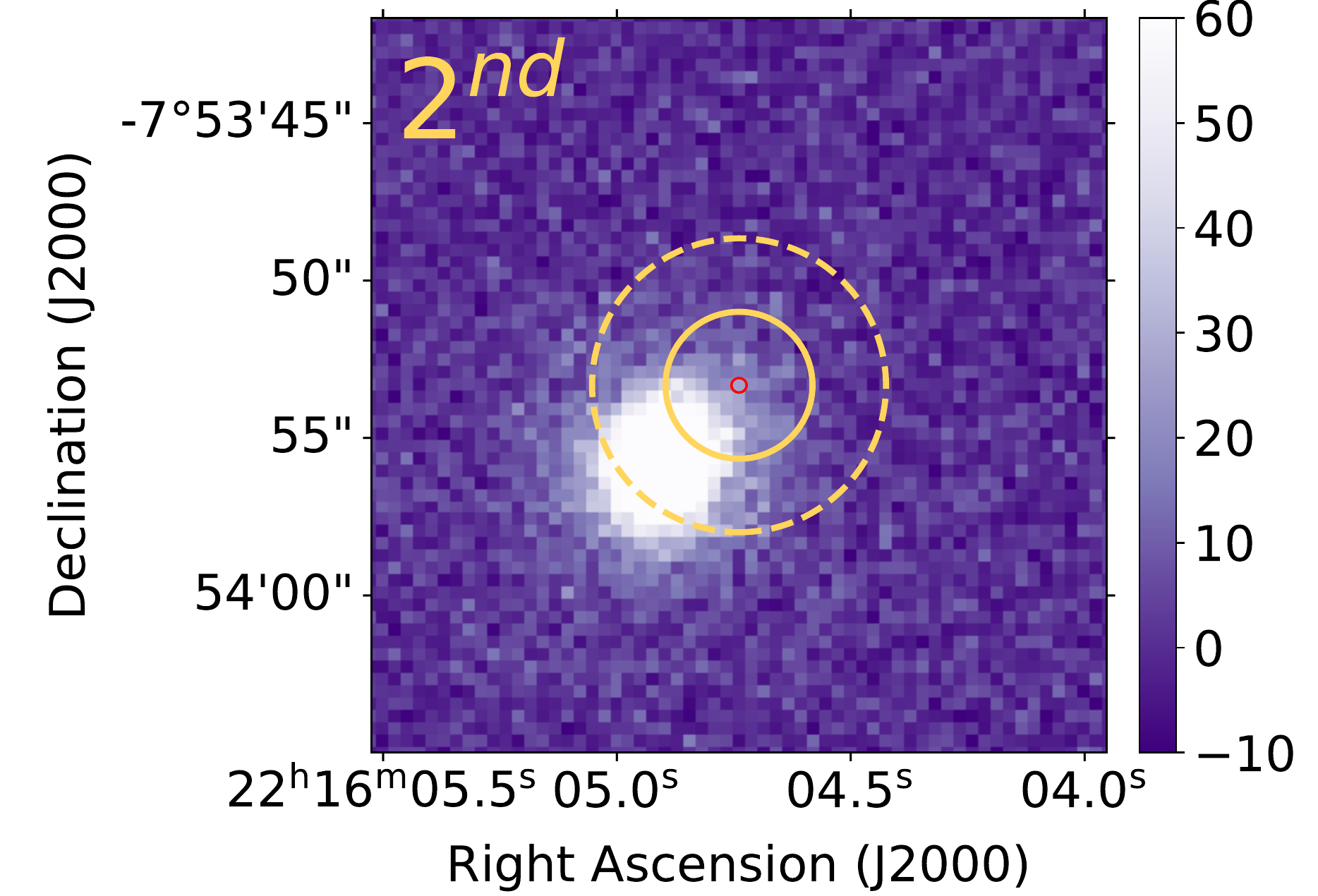}}
    \subfigure[FRB190608 in $3^{rd}$ epoch.]
    {\includegraphics[width=0.3\textwidth]{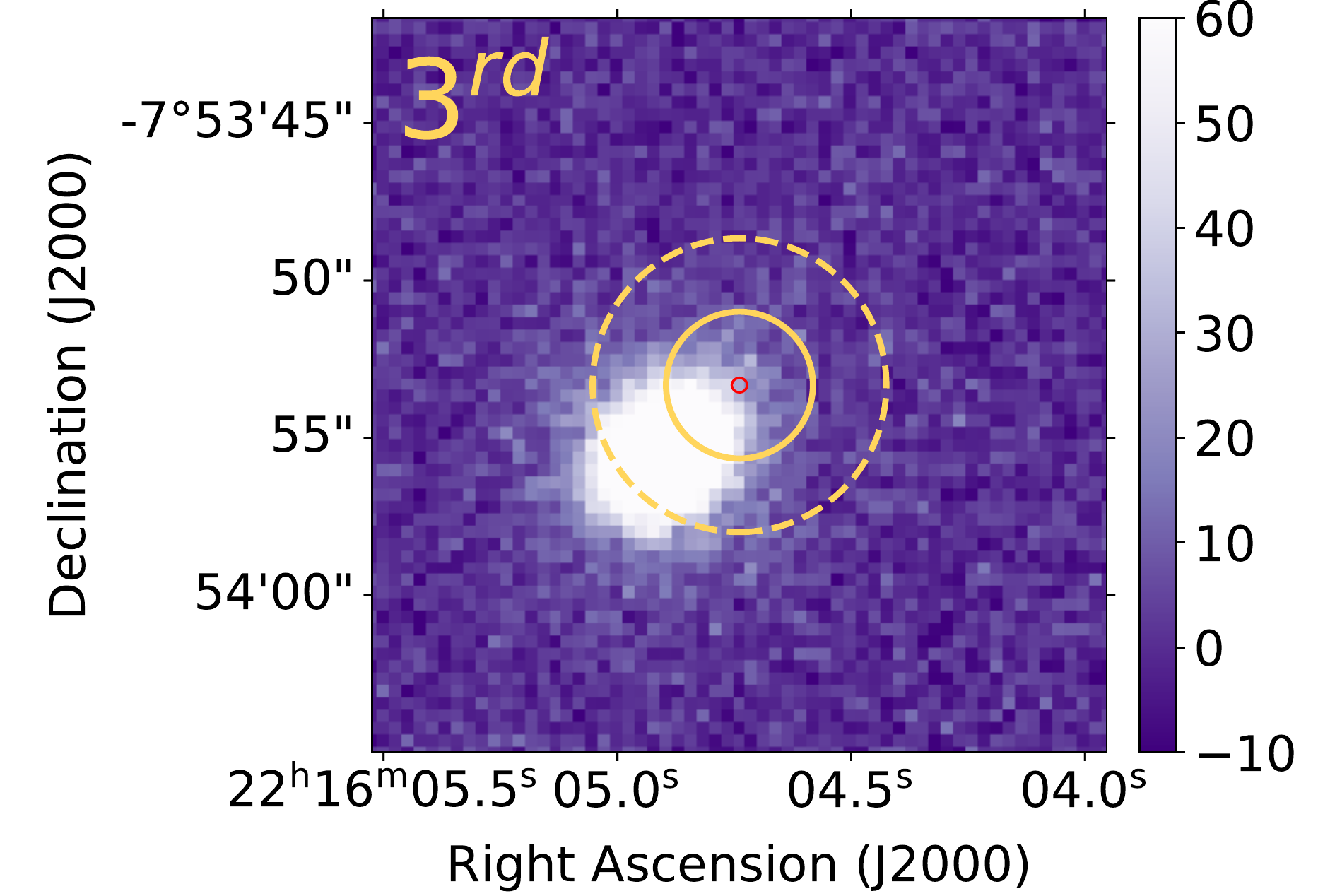}}
    
    \subfigure[Last epoch as template]{
    \includegraphics[width=0.3\textwidth]{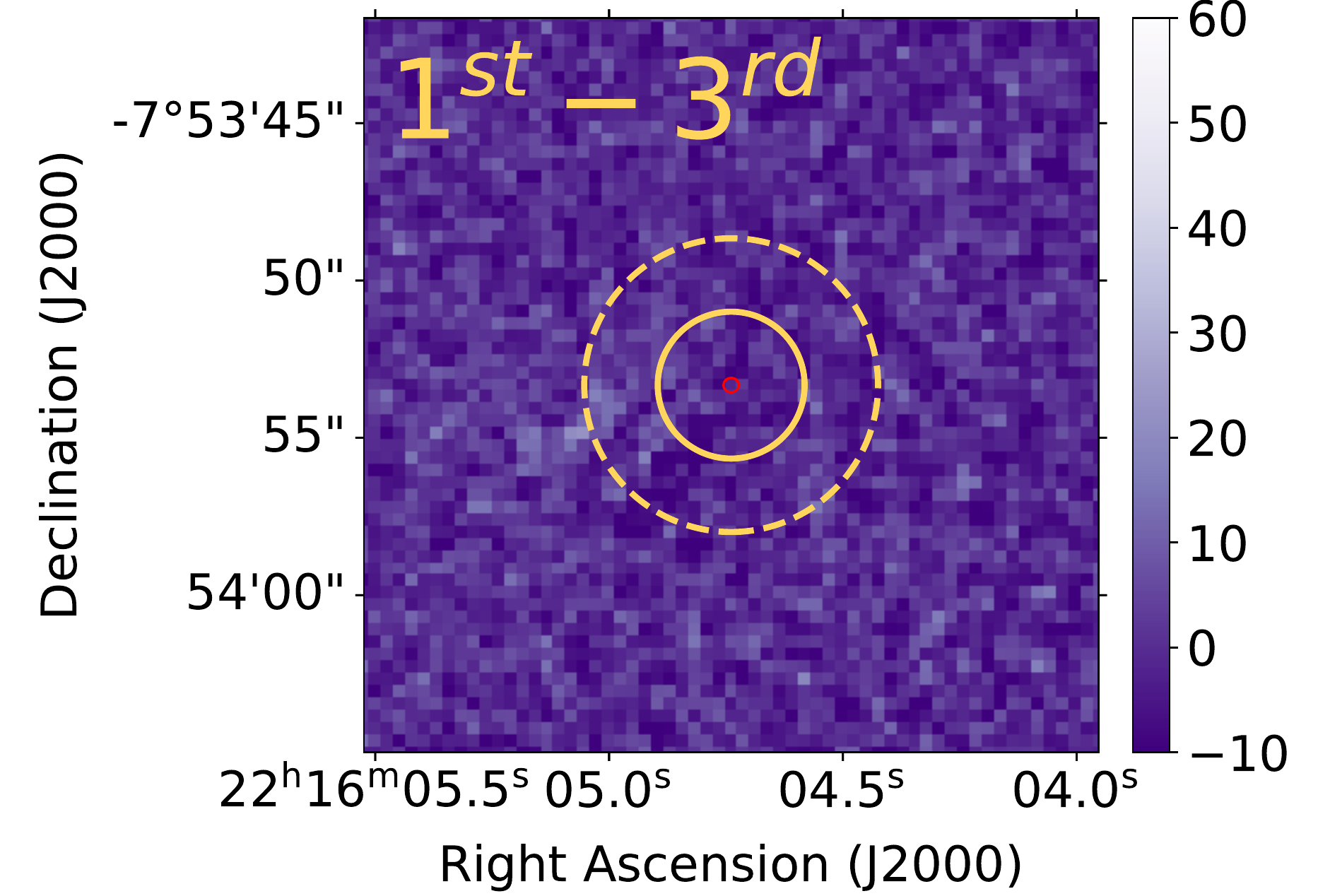}
    \includegraphics[width=0.3\textwidth]{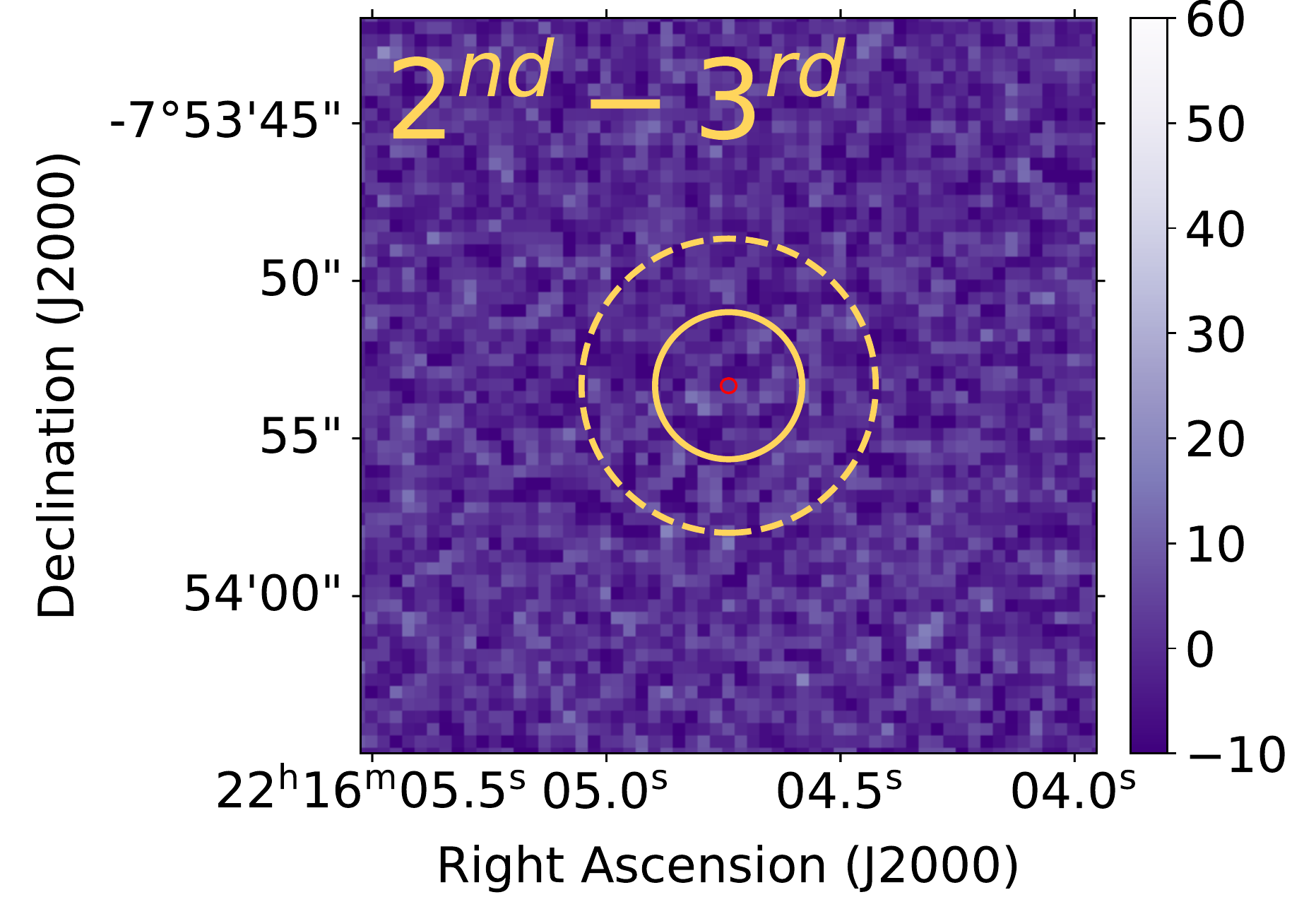}
    }
    
    \subfigure[First epoch as template]{
    \includegraphics[width=0.3\textwidth]{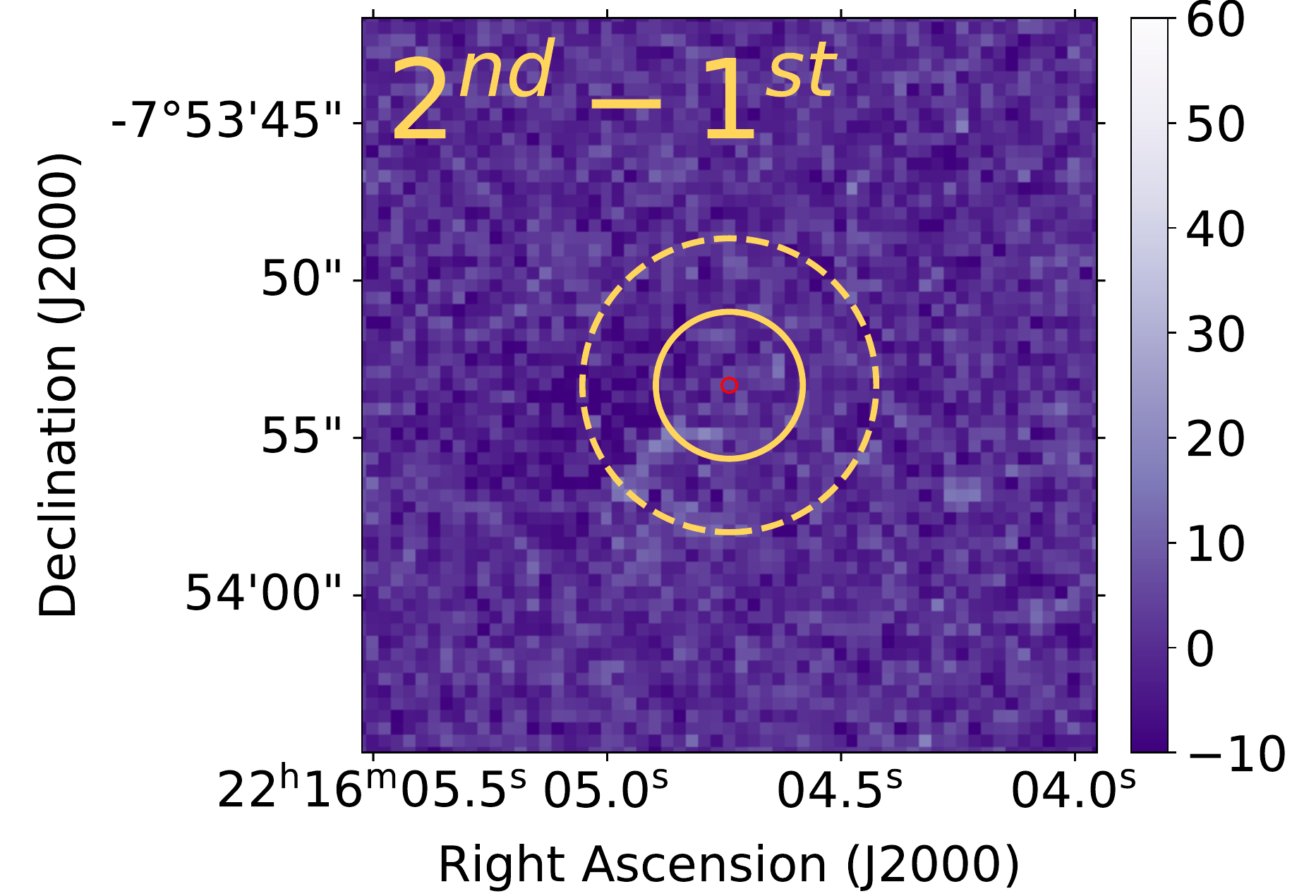}
    \includegraphics[width=0.3\textwidth]{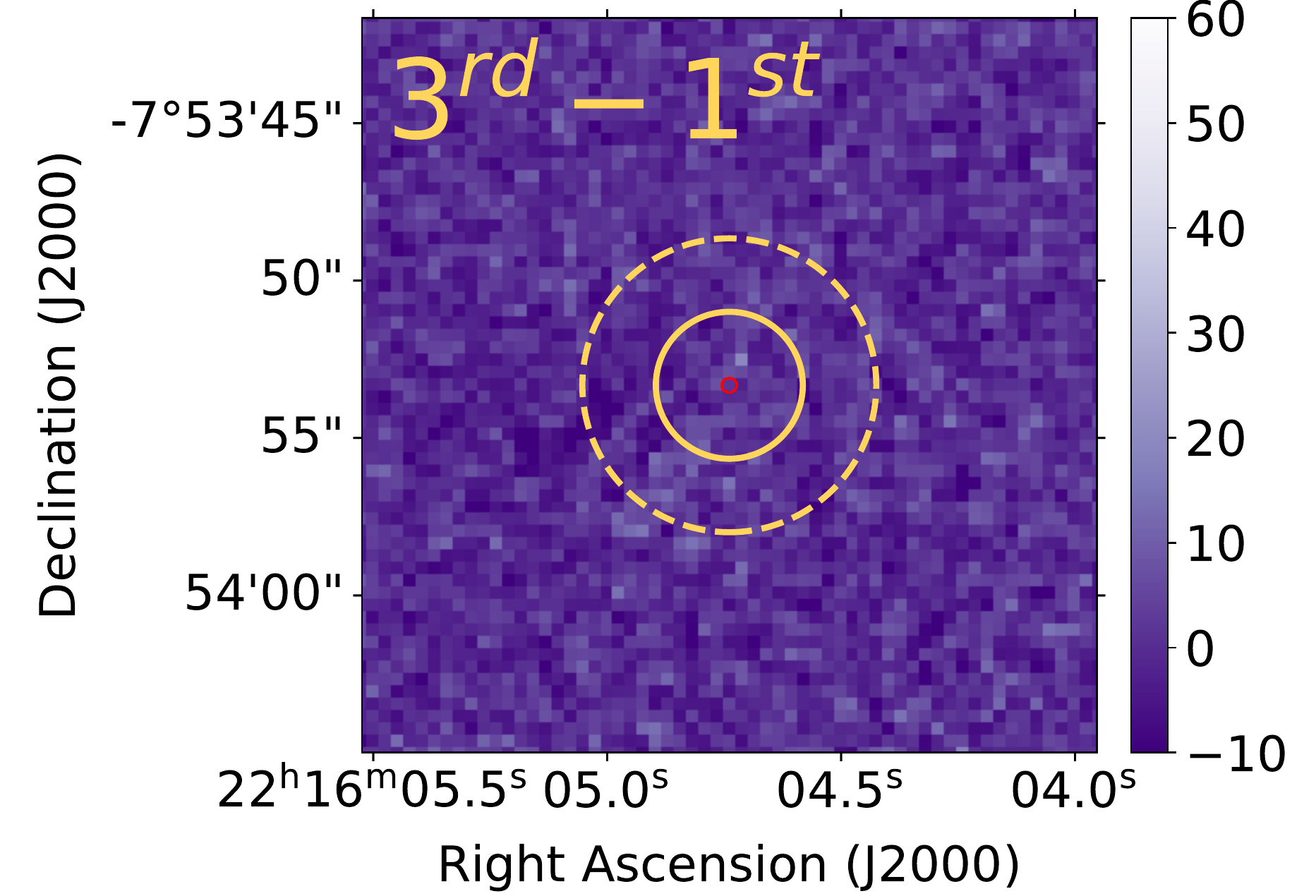}
    }
    
    \caption{Example of difference images for FRB190608. Panels a), b), and c) correspond to the images of each epoch of the FRB190608 observations (three epochs in total for this FRB). Panels d) and e) show the difference images using the image of the last (3) and first (1) epoch as a template, respectively. The solid and dashed circles in all the panels represent the aperture radius of one and two times the FWHM, respectively, as reference, where FWHM = $\sim 2.3 \arcsec$. Our photometric analysis uses an aperture radius of two times the FWHM to set the non-detection limiting magnitudes. The red ellipse represents the uncertainty in the FRB position.}
    \label{differences}
\end{figure*}

\section{Analysis}
\label{sec:analysis}

\subsection{Co-addition and photometry}

We co-added the ten images from each epoch using the SWarp software \citep{swarp}, centering on the FRB coordinates in each field, with an image resampling method, LANCZOS4, and subtracting the background from individual images before resampling.

To obtain the photometry of each FRB field, we used the software Source Extraction and Photometry \citep[SEP;][]{sep, sextractor} on the stacked images. We empirically calibrated the magnitude zero points by performing a crossmatch between the point sources in our fields and the SkyMapper Southern Survey star catalog \citep{skymap}. After subtracting the background estimated by SEP, we calculated fluxes of all point-like sources in our FoV using a circular aperture of $8$ pixel radius ($\sim 3\arcsec$). We did not correct for our fixed aperture, which affects the estimated flux at the order of $5\%$. We estimated the zero point for each stacked frame by crossmatching each point-like, unsaturated source in our imaging to those in the SkyMapper catalog and within a radius of $2\arcsec$.

\subsection{Search for transients}\label{sec:search_for_transients}

We searched for sources of transient optical emission at the precise FRB coordinates for each field. Images of the different epochs in a given FRB field were subtracted using the High Order Transform of PSF And Template Subtraction software \citep{hotpants}. The convolution parameters used are the following: The template was convolved, and we used kernel order 2, background order 1, and ten stamps in the $x$ and $y$ dimensions.

We produced two sets of template-subtracted images. 
In one group, we used the image of the last epoch in a given field as the ``template.'' In the other group, we use the image of the first epoch in a given field as the template. In both cases the template refers to the image that is subtracted from the other. Figure \ref{differences} shows an example of these subtraction groups for FRB190608.

We then ran SEP on the image difference with a circular aperture radius of  twice the full width at half maximum (FWHM) and detection threshold of $1.5\sigma$ in order to look for possible bright optical transients at the position of the eight individual FRBs. No bright optical transients were detected using this method, so we proceeded to set limiting magnitudes.

\subsection{Limiting magnitudes}\label{sec:lower_limits}
    
Since we did not find any optical transients in our data, we set limiting magnitudes to compare such limits with models of possible progenitors.

To estimate the limiting magnitudes, we injected artificial stars (point-like sources) of different apparent magnitudes at the positions of the FRBs in each image, in this manner mocking an unresolved optical transient. We estimated the FWHM of each LCOGT image by taking the average from three well-defined stars in the FoV (reported in the fourth column of Table~\ref{tab:observations}). The artificial star is represented as a 2D Gaussian flux distribution, modeled according to the FWHM of each image.

\begin{figure*}[t]
\centering
  \includegraphics[width=0.325\textwidth]{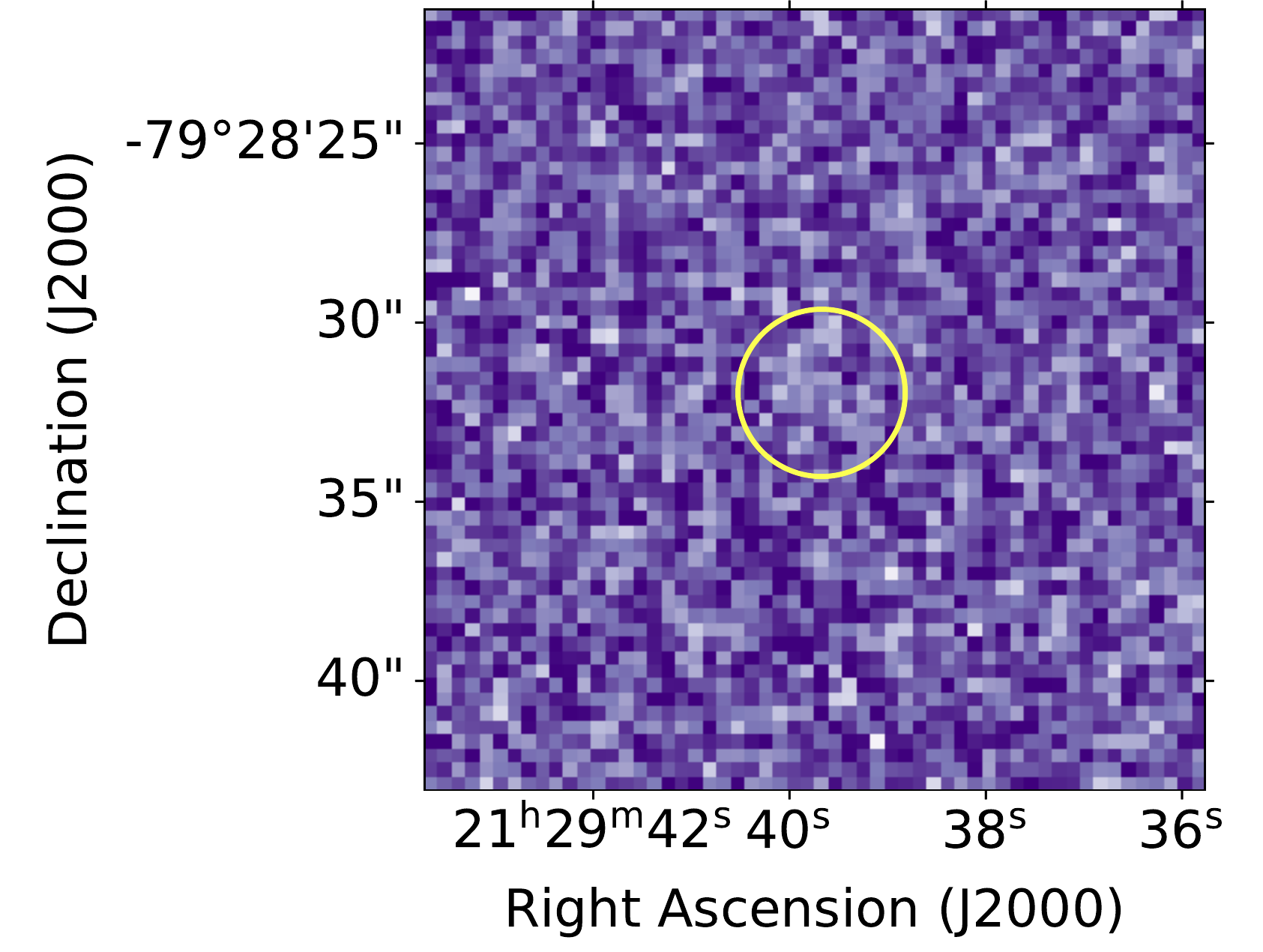}
  \includegraphics[width=0.325\textwidth]{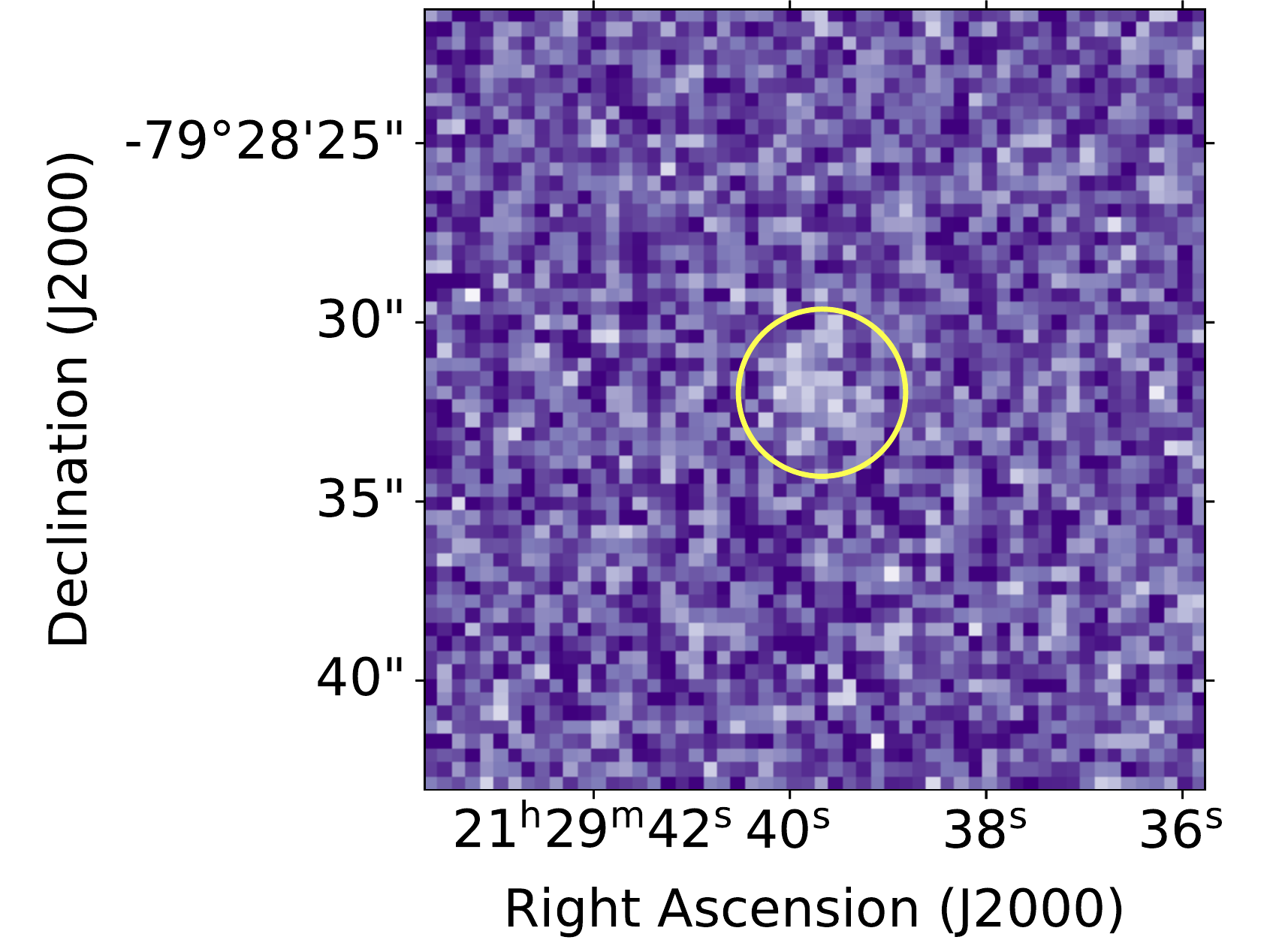}
  \includegraphics[width=0.325\textwidth]{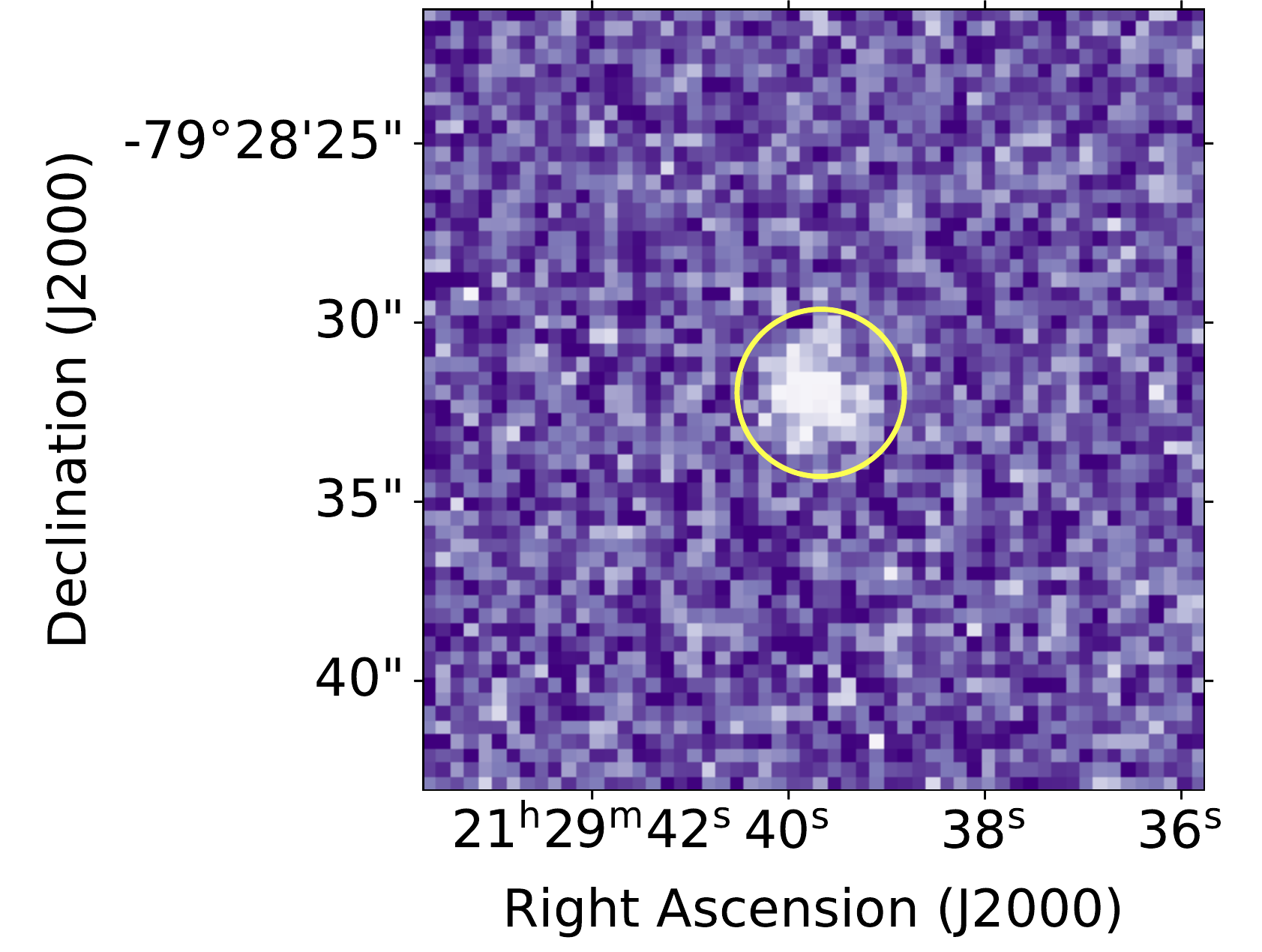}
  \caption{Example of difference images of two epochs of the FRB190102 field, where one of them includes a mock point-like source injected with different magnitudes: a non-detected source (S/N $< 3$; left panel), a limiting magnitude (S/N $=3$; center panel), and a well-detected source (S/N $> 3$; right panel). The pale yellow circle indicates the position where we injected the mock star with an aperture radius of the size of 1 FWHM as reference.}
\label{criteria}
\end{figure*}

After adding the artificial star to an individual image, we template-subtracted it (using either the first or last epoch; see Sect.~\ref{sec:search_for_transients}). Then, we proceeded with the same automatic detection described above (see Sect.~\ref{sec:search_for_transients}) and recorded the resulting signal-to-noise ratio (S/N) of the source recovered, calculated as the ratio between the recovered flux and its recovered flux error at the position of the mock transient (i.e., the position of the FRB). We repeated this process for artificial stars of different apparent magnitudes, ranging from $17$ to $25$\,mag and using $0.1$\,mag steps. 
    
When the S/N of the recovered star was greater than 3, we considered that a detection, and vice versa
(see Fig.~\ref{criteria} for an illustration of this criterion). The recovered apparent magnitude was calculated through the flux, using a circular aperture radius of twice the FWHM. The limiting apparent magnitude in a given epoch was set as the fainter recovered magnitude from the difference image with S/N $>3$. 

Since we have the redshift of the host galaxies of FRBs (i.e., the redshift of the FRB as well), we obtained the absolute magnitudes using the distance modulus and taking into account Galactic extinction (see Table~\ref{tab:epochs}). All the limiting magnitudes for our different epochs are presented in Table \ref{tab:mags}.

\begin{table*}[t]
\caption{Limiting magnitudes for each epoch.}
\label{tab:mags}      
\centering          
\begin{tabular}{c c c c c c c c c c}     
\hline\hline       
FRB & \multicolumn{8}{c}{Limiting Magnitude} \\ \hline
& \multicolumn{4}{c}{Last epoch as template} & \multicolumn{4}{c}{First epoch as template}\\
\hline
& \multicolumn{2}{c}{1$^{st}$ epoch} & \multicolumn{2}{c}{2$^{nd}$ epoch} & \multicolumn{2}{c}{2$^{nd}$ epoch} & \multicolumn{2}{c}{3$^{rd}$ epoch}\\
& Apparent & Absolute & Apparent & Absolute & Apparent & Absolute & Apparent & Absolute \\
& (mag) & (mag) & (mag) & (mag) & (mag) & (mag) & (mag) & (mag) \\
\hline                    
    FRB180924 & 22.5 & -18.7 & - & - & 21.9 & -19.3 & - & - \\
    FRB181112 & 21.4 & -20.8 & - & - & 21.2 & -21.0 & - & - \\
    FRB190102 & 21.8 & -19.6 & - & - & 22.3 & -19.1 & - & - \\
    FRB190608 & 21.7 & -17.1 & 20.6 & -18.2 & 21.6 & -17.2 & 21.7 & -17.1 \\
    FRB190611 & 21.1 & -21.0 & 21.4 & -20.7 & 21.9 & -20.2 & 22.0 & -20.1 \\
    FRB190711 & 22.1 & -20.6 & 20.6 & -22.1 & 20.5 & -22.2 & 21.6 & -21.1 \\
    FRB190714 & 21.0 & -19.5 & 20.4 & -20.1 & 20.8 & -19.7 & 22.1 & -18.4 \\
    FRB191001 & 23.7 & -16.7 & 21.8 & -18.6 & 22.0  & -18.4 & 21.2 & -19.2 \\
\hline            
\label{ap_mag}
\end{tabular}
\end{table*}

\section{Results}
    \label{sec:results}

Figure \ref{ab_mag_plot} shows the limiting absolute magnitudes of the eight well-localized FRBs studied here at different epochs.\ The filled and empty triangles represent those that use the last and first epoch as a template, respectively (see Sect. \ref{sec:search_for_transients}).

Although one could expect the limits derived using the first or last epochs as templates to be the same, this is not always the case because the template subtraction process depends on the quality of both images involved in the process (e.g., background, point-spread function). Indeed, we observe a significant difference (although $\leq 1$\,mag) in the limits inferred for the second epochs of FRB190608, FRB190611, and FRB190714; 
these differences come mostly from the fact that the first and last epochs were observed with large differences in seeing conditions and/or moonlight illumination (see Table \ref{tab:observations}).

\subsection{Comparison of limiting magnitudes with SN light curves}
\label{subsec:comparison_SN}
    
In the left panel of Fig.~\ref{ab_mag_plot} we include different light curves of bright optical transients to compare with the limits inferred above, assuming the FRB coincides with the triggering of the transient explosion. In particular, we focus the comparison on different types of bright SNe, obtained from the Open Supernova Catalog \citep{SNcat}. The catalog provides the data points of each light curve in apparent magnitudes, which we converted to absolute magnitudes by adding the distance modulus (given the redshift of the SNe provided by the catalog), taking into account Galactic extinction but not the intrinsic extinction of the host galaxy nor any $K$ correction.

For this comparison we considered SLSN, GRB-SN, and SN~Ia types. In order to have a representative range of SN light curves, we chose the brightest and the faintest of each type available in the catalog: For SLSNe these were, respectively, SN2015bn \citep{SLSN} and SN2010md \citep{ptf10hgi}; for GRB-SN they were  SN1998bw \citep{sn1998bw} and SN2010bh \citep{sn2010bh}; and for SN~Ia they were LSQ12gdj \citep{lsq} and SN2009F \citep{krisciunas17}. We cut the longest sampled light curve in a given type to match the extent of the shortest. We then interpolated the corresponding data points to produce a ``light-curve band'' for each type, as shown in the left panel of Fig.~\ref{ab_mag_plot}. As reference, the light curve of the kilonova AT170817 \citep{kilonova} connected with the gravitational wave emission GW170817 \citep{abbott17} is also shown.

Assuming that the emission of the FRB coincides with the day of the explosion of the transients, we can use Fig.~\ref{ab_mag_plot} to test possible associations. In particular, we can rule out SLSNe as a possible progenitor model for all FRBs since we have four limiting magnitudes under the light curve of the faintest SLSN cataloged. That is, if there had been an optical transient associated with an SLSN, we would have seen it in our LCOGT data at a confidence of $\sim99.9\%$ (see Sect. \ref{quantification} for details). In contrast, for SNe~Ia and GRB-SNe we can only rule out the bright end of the class because normal and under-luminous SNe~Ia lie below our detection limits. Similarly, our data are not deep enough to rule out kilonovae as possible progenitors of FRBs.

\subsection{Quantification for ruling out SLSNe}
    \label{quantification}
    
Thus far, we have set the detection threshold as a constant value of $1.5\sigma$ in the difference image. This confidence interval gives us an $86.6\%$ probability of finding the FRB between $x-1.5\sigma < x < x+1.5\sigma$, but since they are limiting magnitudes, we integrate the probability distribution above $x+1.5\sigma$, which results in $93.3\%$. That is, we have a $93.3\%$ probability that the putative optical counterpart of the FRB is below its limiting magnitude and thus a $6.7\%$ probability that it is above the reported limiting magnitude. Because we have multiple independent limits below the faintest SLSN, we can generate more stringent limits by combining these observations.
    
Although subtracting two images from two different epochs 
in reverse order yields two limiting magnitudes for some epochs, it is important to note that these measurements are not independent. Thus, in order to combine limiting magnitudes, we only take the most stringent one in a given epoch for a given FRB field that lies below the faintest SLSN curve. This gives a total of four independent measurements.

Since we have four independent limiting magnitudes under the faintest SLSN light curve, the probability of being above the four limiting magnitudes is reduced to $\sim0.002\%$. In other words, under the assumption that all FRB emissions are simultaneous with the SLSN explosion, we can rule out SLSNe as a progenitor of FRBs with a confidence level (c.l.) of $\sim99.998\%$. On a similar basis, we can rule out the brightest subtypes of SNe~Ia and GRB-SNe at a similar confidence.

We note that these estimations do not consider systematic effects, such as different arrival times for the FRB. However, we can relax the assumption of explosion-FRB simultaneity. For this we considered time delays for the FRB that guarantee a minimum number of magnitude limits remaining under the faintest SLSN light curve (i.e., equivalent to continuously moving the SLSN light-curve band to the left in Fig.~\ref{ab_mag_plot} and comparing it with the limits). If we take a minimum of two limiting magnitudes (corresponding to a $\sim99.6\%$ c.l.), we can still rule out an FRB-SLSN association for scenarios where the FRB arrival is up to $170$\,days after the SLSN explosion.

These estimations also do not consider systematic uncertainties in the object luminosity due to host galaxy extinction. Nevertheless, our limits do not change significantly, even with $r=1$\,mag of intrinsic extinction. In such a case, we obtain three limiting magnitudes of FRBs under the SLSN light-curve band instead of four. However, if we include the limiting magnitude associated with the first epoch of FRB191001 (which is well below $1$\,mag of the extrapolation of the SLSN to $<1$\,day earlier; see Fig.~\ref{ab_mag_plot}), our previous significance remains the same.

\subsection{Comparison of limiting magnitudes with TDE light curves}

\begin{figure*}[t]
\centering
  \includegraphics[width=0.615\textwidth]{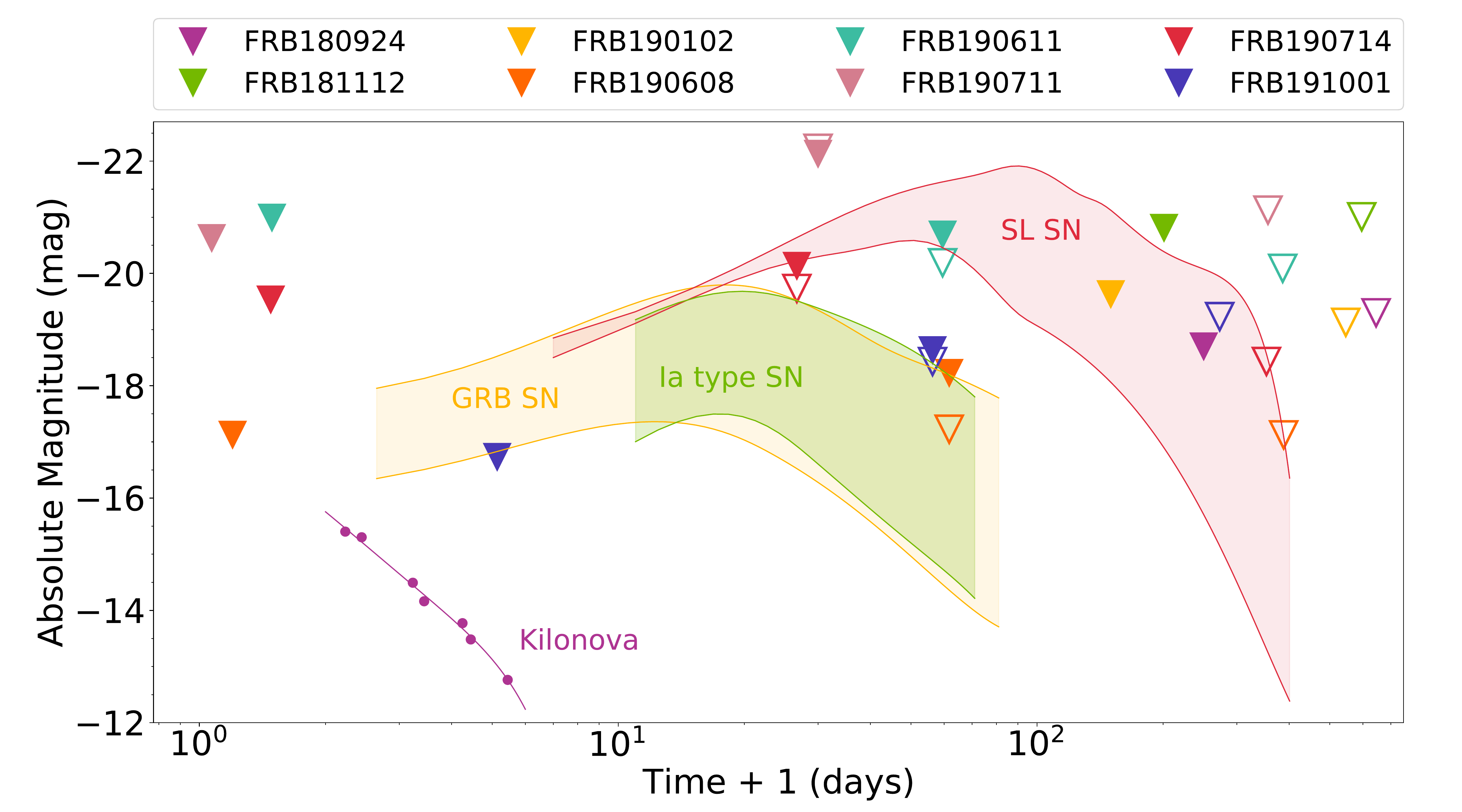}
  \includegraphics[width=0.375\textwidth]{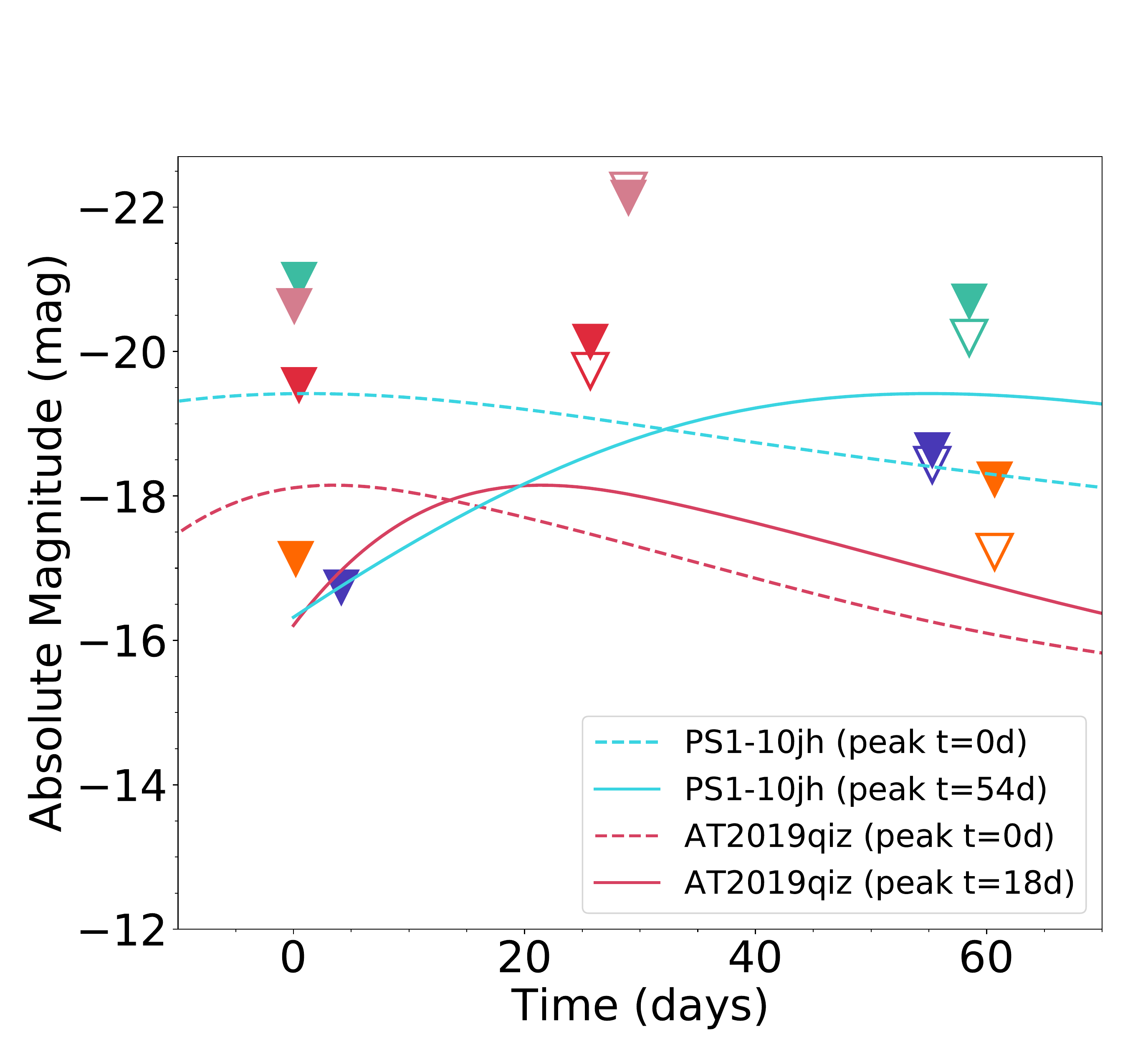}
  \caption{Limiting absolute magnitudes at different epochs obtained for FRB positions (triangles) and light curves of different bright optical transients (colored lines and regions). Filled and open triangles correspond to the limit that uses the last and the first epoch as a template, respectively. {\it Left panel:} Light-curve bands (see Sect.~\ref{subsec:comparison_SN}) for SLSNe (red), GRB-SNe (yellow), and SNe~Ia (green). The light curve of the optical counterpart to a kilonova (GW170817) is also shown (purple). For the limiting magnitudes, the $x$ axis corresponds to the time since the FRB signal was received on Earth; for the light curves, the $x$ axis corresponds to the time since explosion. We emphasize that the $x$ axis is shown on a logarithmic scale ($+ 1$\,day for convenience) and that the $y$ axis is on the astronomical magnitude scale, such that brighter objects are at the top. {\it Right panel:} Representative TDE light curves (PS1-10jh in cyan, AT2019qiz in pink) with luminosity peak times coinciding with the FRB arrival time (dashed lines) and with $t=54$ and $t=18$\,days (solid lines) after the FRB arrival time. Here the $x$ axis is presented in linear scale.}
  \label{ab_mag_plot}
\end{figure*}

In the right panel of Fig.~\ref{ab_mag_plot} we have focused on the limiting magnitudes within the first few days after the arrival of the FRB. As reference, we have added light curves of two TDEs, AT2019qiz \citep{TDEAT} and PS1-10jh \citep{TDEPS1}, obtained from the Open TDE Catalog.\footnote{http://TDE.space}

In contrast to the case of SNe that have a well-defined time of explosion, for TDEs it is not possible to define an obvious initial time. For this reason, in the following we consider two broad possibilities: 
(1) the FRB is emitted when the TDE reaches its peak luminosity (dashed lines in the right panel of Fig.~\ref{ab_mag_plot}), and (2) the FRB is emitted $54$~days or $18$~days before the TDE reaches its peak luminosity (solid lines in the right panel of Fig.~\ref{ab_mag_plot}). The second possibility is motivated by being the time between the first cataloged data point and the peak luminosity of each TDE. Although arbitrary, this choice provides a range of possibilities for a putative FRB-TDE association.

Taking into account possibility (1), we have two independent limiting magnitudes under the light curve of the faintest TDE. Following the reasoning presented in Sect.~\ref{quantification}, in this case we can rule out TDEs as possible progenitors of FRBs with a $\sim99.6\%$ c.l. Now focusing on possibility (2), we can only rule out the brightest TDE, which also has two independent limiting magnitudes under the light curve (i.e., $\sim99.6\%$ c.l.).

\subsection{Dearth of prompt high-energy emission from GRB-SNe}

We can put further constraints on the association of FRBs with GRB-SNe by quantifying the apparent lack of simultaneous high-energy prompt emission from the radio bursts. Since the prompt gamma-ray burst (GRB) emission is highly beamed \citep{Gehrels09}, only a fraction of FRB-emitting sources (in the case that $\gamma$-ray emission is directly associated with them) will have a detectable precursor GRB. This can be expressed approximately as $N_{\rm tot} = N_{\rm visible}/(1-\cos(\theta_{\rm jet}))$, such that for average jet opening angles, $\theta_{\rm jet} < 10^\circ$ \citep{Frail01,Bloom03,Guetta05}, only $\approx 1-2\%$ of GRBs will be detected. Convolved with the typical fraction of GRBs detectable by, for example, {\it Swift}/BAT at any given time \citep{Lien14}, the probability of detecting a GRB associated with an FRB is $\lesssim 1\%$, assuming that the FRB emission is isotropic (or at least not beamed in a similar way as the GRB). We thus cannot rule out that prompt GRBs can be associated with FRBs given the still small number of well-localized FRBs \citep[see also][]{palaniswamy14,xi17,martone19,guidorzi20}. However, considering the wide FoV and precise timing of GRB searches, it may be feasible to use the large number of FRBs with poorer angular resolution \citep[e.g.,][]{cat_chime} to conduct a similar statistical assessment.

\section{Conclusions}
\label{sec:conclusions}

The search for progenitors of FRBs has become a great challenge for the astronomical community over the last few years. Unlike the works carried out by \cite{optical}, \cite{op_follow_up}, and \cite{optical2}, who were looking for an optical counterpart associated only with a particular FRB, in this work we have a larger data sample, analyzing FoVs toward eight different FRBs.

Although we have not found an optical transient that solves the mystery of the progenitors of FRBs, we have ruled out SLSNe being a dominant channel, at least over the timescales probed here ($\approx 170$\,days) and for galaxy host extinctions of $\lesssim 1$ magnitude. Super-luminous SNe are rare explosions from poorly understood astrophysical phenomena associated with the ending lives of massive stars. They emit approximately $100$ times more energy than typical SNe \citep{jerkstrand}. With the data and the analysis obtained in this work, we can rule out the association of SLSNe with FRBs with a c.l. of $\sim99.99\%$, assuming the FRB emission coincides with the SLSN explosion. However, this does not rule out the possibility that the FRBs come from a particular object, such as an NS, which is surrounded by an extreme environment, such as an SLSN.

For SNe~Ia and GRB-SNe, we can only rule out the most luminous of each class as the faintest ones are too weak to be detected in our LCOGT data given our inferred limits and distances to the FRB hosts. Furthermore, in contrast to the work done by \cite{marnoch}, we have data from $\sim1$ day after the arrival of the FRBs, which allowed us to rule out the brightest TDEs on the basis of photometry. Finally, we cannot rule out a kilonova.

\begin{acknowledgements}

We thank the anonymous referee for her/his constructive comments and suggestions which have improved the quality of this manuscript. This work makes use of observations from the Las Cumbres Observatory global telescope network (LCOGT), obtained as part of programs CN2019A-39/CLN2019A-002, CN2019B-93/CLN2019B-001 and CN2020A-82/CLN2020A-001.
C.N. and N.T. acknowledge support by FONDECYT grant 11191217. G.P. acknowledges support by ANID - Millennium Science Initiative - ICN12$\_$009. K.E.H. acknowledges support by a Postdoctoral Fellowship Grant (217690--051) from The Icelandic Research Fund. A.T.D. is the recipient of an ARC Future Fellowship (FT150100415). R.M.S. acknowledges support through ARC Future Fellowship FT190100155. L.M. acknowledges the receipt of an MQRES scholarship from Macquarie University. The Fast and Fortunate for FRB Follow-up team acknowledges support from NSF grants AST-1911140 and AST-1910471. 

\end{acknowledgements}

\bibliographystyle{aa}
\bibliography{main.bib}

\end{document}